\documentclass[twocolumn,pra,showpacs]{revtex4}
\usepackage{amsmath}
\usepackage{graphicx}

\newcommand{\eq}{\begin{equation}}
\newcommand{\fine}{\end{equation}}

\begin{document}

\title{Decoherence, Einselection and Classicality of a Macroscopic Quantum
Superposition generated by Quantum Cloning}
\author{Francesco De Martini$^{1,2}$, Fabio Sciarrino $^{1}$, and Nicol\`{o}
Spagnolo$^{1}$}

\address{$^{1}$Dipartimento di Fisica dell'Universit\'{a} ''La Sapienza'' 
and Consorzio Nazionale Interuniversitario per le Scienze Fisiche della Materia, Roma, 00185 Italy\\
$^{2}$ Accademia Nazionale dei Lincei, via della Lungara 10, I-00165 Roma, Italy }

\begin{abstract}
The high resilience to de-coherence shown by a recently discovered
Macroscopic Quantum Superposition (MQS)\ generated by a quantum injected
optical parametric amplifier (QI-OPA) and involving a number of photons in
excess of $5\times 10^{4}$ motivates the present theoretical and numerical
investigation. The results are analyzed in comparison with the properties of
the MQS\ based on $\left| \alpha \right\rangle $ and NOON\ states, in the
perspective of the comprehensive theory of the subject by W.H.Zurek. In that
perspective the concepts of ''pointer state'', ''einselection'' are applied
to the new scheme.
\end{abstract}

\maketitle

\section{Introduction}

The short handwritten note by Einstein on the back of a greetings card sent
to Max Born on the first of January 1954 may be taken as the conceptual
framework of the present work:\ ''\textit{if }$\varphi _{1}$\textit{and }$%
\varphi _{2}$\textit{\ are two solutions of the same Schr\"{o}dinger
equation, }$\varphi =\varphi _{1}+\varphi _{2}$\textit{\ is another solution
of the same equation equally able to represent a possible situation. If
however we are dealing with a ''macrosystem'' and }$\varphi _{1}$\textit{and 
}$\varphi _{2}$\textit{\ are ''narrow'' respect to the macrocoordinates in
the vast majority of cases }$\varphi $\textit{\ cannot be ''narrow''.
Narrowness respect to the macrocoordinates [}i.e. macro-localization\textit{%
] is a property not only independent of the principles of quantum mechanics
but also incompatible with them }'' \cite{Eins65}.

As we can see since the early decades of Quantum Mechanics the
counter-intuitive properties associated with the superposition state of
macroscopic objects and the problem concerning the ''classicality''\ of
quantum macrostates were the object of an intense debate epitomized in 1935
by the celebrated ''\textit{Schr\"{o}dinger Cat\ paradox}'' \cite%
{Eins35,Schr35}. In particular, the actual feasibility of such quantum
object has always been tied to the alleged infinitely short persistence of
its quantum coherence, i.e. of its overwhelmingly rapid ''decoherence''. In
modern times the latter property, establishing a rapid merging of the
quantum rules of microscopic systems into classical dynamics, has been
interpreted as a consequence of the entanglement between the macroscopic
quantum system with the environment \cite{Niel00,Zure}. By tracing over the
environmental variables in the final calculations, generally the pure
quantum state decays irreversibly towards a probabilistic classical mixture 
\cite{Dur02}. Recently, the general interest for decoherence has received a
renewed interest in the framework of quantum information theory where it
plays a fundamental detrimental role\ since it conflicts with the
experimental realization of the quantum computer or of any quantum device
bearing any relevant complexity \cite{Gori07}. In this respect a large
experimental effort has been devoted recently to the implementation of
Macroscopic (i.e. many-particle) Quantum Superpositions states (MQS),
adopting photons, atoms and electrons in superconducting devices. Particular
attention has been devoted to the realization of the MQS involving
''coherent states''\ of light, which exhibits interesting and elegant Wigner
function representations \cite{Schl01}. The most notable results of this
experimental effort have been reached with atoms interacting with microwave
fields trapped inside a cavity \cite{Brun92,Raim01} or for freely
propagating fields \cite{Ourj06}.\ However, in spite of the long lasting
efforts spent in these endeavors, in these realizations the MQS\ has always
proved to be so fragile that even the loss of a single particle\ was found
to be able to spoil any possibility of a direct observation of its quantum
properties. Precisely on the basis of these negative results in many
scientific communities (and also within some influential editorial teams)
grew the opinion that the ''Schr\"{o}dinger Cat''\ is indeed an ill defined
and then avoidable concept since\ it fundamentally lacks of any directly
observable property \cite{Dur02}.

In spite of these conclusions, very recently a new kind of MQS\ involving a
number of particles $N$ in excess of $5\times 10^{4}$ has been realized
allowing the direct observation of entanglement between a microscopic
(Micro-) and a macroscopic (Macro-) photonic state and showing a very high
resilience to decoherence by coupling with environment \cite{DeMa08}.
Precisely, the MQS was generated by a quantum-injected optical parametric
amplifier (QI-OPA) seeded by a single-photon belonging to an EPR entangled
pair. We emphasize here that the reported QI-OPA can be considered for the
present purpose as a paradigmatic system consisting of the simplest
realizable \textquotedblright \textit{optimal\ phase-covariant} \textit{%
quantum cloning machine\textquotedblright\ }\cite{Scia05,Scia07}. Indeed,
precisely the process of \ \textquotedblright \textit{quantum cloning}%
\textquotedblright\ was there responsible for the transfer of the
entanglement and the superposition properties of a pure single-particle
qubit into a multiparticle MQS. In other words, the QI-OPA\ encoded
\textquotedblright optimally\textquotedblright\ into a Macro-state the
information associated with the input Microstate, a \textit{seed} qubit \cite%
{DeMa98,DeMa05,DeMa05b,Naga07,Ricc05}. By this device, that includes an
Orthogonality Filter (O-Filter) for enhanced state discrimination, the
Micro-Macro state non-separability was successfully tested and the
Micro-Macro violation of the Bell'inequalities for Spin-1 excitations was
attained \cite{DeMa08,DeMa08b}. In view of this peculiar, striking behavior,
we felt that a careful analysis\ of the decoherence of this novel MQS\
device was necessary. The present approach to decoherence will be cast
within the\ useful framework developed in the past by Wojciech H.Zurek\cite%
{Zure03}. Accordingly, the systems we shall consider will be characterized
according to the concepts of \ \textquotedblright \textit{information flow}%
\textquotedblright ,\ \textquotedblright \textit{pointer states}%
\textquotedblright , \textquotedblright \textit{einselection}%
\textquotedblright . Precisely, according to Zurek's definition, the
"pointer states" of any detector, or of any open quantum system, consist of
a preferred basis that is selected , i.e. "einselected" by the
characteristic detector - environment interaction. When expressed in terms
of these states, the correct density matrix for the detector - system
combination can be obtained by standard Schr\"{o}dinger equation theory
without having to appeal to the von Neumann's nonunitary "reduction"
process. These privileged pointer states
represent the natural behavior of the detector's "pointer" expressing the
physical outcome of any measurement process. The present work is intended to
provide a new insight in the elusive, fundamental problem of decoherence
and, at last, it will provide a sensible reply to the argument expressed in
the Einstein letter to Born. Hopefully, it may also lead to a revision of
several prejudices about MQS.

Let's summarize the content of the present paper. In Section II, we
introduce the criteria adopted to characterize the resilience to decoherence
of \ any MQS while, in order to test the validity of our approach, these
criteria are applied in Section III to the case of the "coherent" Glauber's
- state MQS. We obtain an universal function which is in agreement to the
expected exponential decrease of coherence, induced by losses, peculiar of
this class of states. The central point of the paper is addressed in Section
IV where a thourough computer analysis of the decoherence affecting the MQS\
generated by the process of \emph{optimal} phase-covariant quantum cloning
of a single-photon state. The applicaton of the criteria here introduced
show the very high resilience to losses of this MQS. Finally, by the
conclusive Section V the scope of the work is considered on a broader
perspective involving some relevant, basic quantum mechanical issues.

\begin{figure}[t]
\centering
\includegraphics[width=0.5\textwidth]{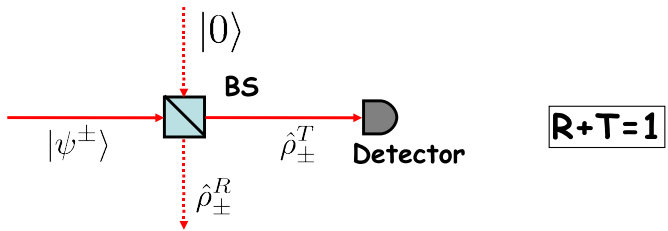}
\caption{(Color online) Schematization of the decoherence model by a linear beam-splitter
of transmittivity T.}
\label{fig:fig1}
\end{figure}

\section{Definitions and Criteria}

In this Section we introduce the method we intend to adopt in order to
provide a consistent investigation of the "resilience to decoherence" of any
MQS. Precisely, we introduce first the criteria, based on the concept of \
"distinguishability" between two orthogonal quantum states and the related
"degree of coherence" of a MQS involving the same states. The parameter here
introduced, the "Bures distance" is assumed as the "merit figure" expressing
the persistance of quantum effects when decoherence process progressively
randomizes the relative phase between the two components of \ the MQS.
Finally, we describe the model adopted to simulate the transmission of the
analyzed field over a "lossy channel".

\textbf{Criteria for macroscopic superposition.} In order to distinguish
between two different quantum states, let us introduce the definition of \
''distance'' $D$ in the Hilbert space, i.e. a parameter which expresses
quantitatively the overlap of two generic states $\widehat{\rho }$ and $%
\widehat{\sigma }$ via the\ ''fidelity''\ $\mathcal{F}(\widehat{\rho },%
\widehat{\sigma })=\mathrm{Tr}\left( \sqrt{\widehat{\rho }^{\frac{1}{2}}%
\widehat{\sigma }\widehat{\rho }^{\frac{1}{2}}}\right) \ $being $0\leq 
\mathcal{F}\leq 1$, where $\mathcal{F}=1$ for $\widehat{\rho }=\widehat{%
\sigma }$, and $\mathcal{F}=0$ for orthogonal states \cite{Jozs94}. This
quantity is adopted to define the ''Bures distance'', a metric in the state
space: $D(\widehat{\rho },\widehat{\sigma })=\sqrt{1-\mathcal{F}(\widehat{%
\rho },\widehat{\sigma })}$ \ \cite{Bure69,Hubn92}.

\textbf{Distinguishability, MQS\ ''Visibility''.} Let's characterize two
macroscopic states $|\phi _{1}\rangle $ and $|\phi _{2}\rangle $ and the
corresponding MQS's: $|\phi ^{\pm }\rangle =\frac{\mathcal{N}_{\pm }}{\sqrt{2%
}}\left( |\phi _{1}\rangle \pm |\phi _{2}\rangle \right) $ by adopting two
criteria. \textbf{I)} The ''\textit{Distinguishability''} between $|\phi
_{1}\rangle $ and $|\phi _{2}\rangle $ expressed by: $D\left( |\phi
_{1}\rangle ,|\phi _{2}\rangle \right) $. \textbf{II)} The ''\textit{%
Visibility''}, i.e. ''degree of orthogonality''\ of two MQS's $|\phi ^{\pm
}\rangle $ expressed by $D\left( |\phi ^{+}\rangle ,|\phi ^{-}\rangle
\right) .$ Indeed, the value of the MQS\ visibility depends exclusively on
the relative phase of the component states:$\ |\phi _{1}\rangle $ and $|\phi
_{2}\rangle $. The parameter $D$ expresses the ability of an observer to
discriminate between two initially orthogonal states, $D\left( |\phi
^{+}\rangle ,|\phi ^{-}\rangle \right) =1$, after propagation in a\textit{\
''lossy channel''} where the relative phase of $|\phi _{1}\rangle $ and $%
|\phi _{2}\rangle $ progressively randomizes leading to a fully mixed state: 
$D\left( |\phi ^{+}\rangle ,|\phi ^{-}\rangle \right) = 0$.

As we shall see later in the paper, the physical interpretation of $D\left(
|\phi ^{+}\rangle ,|\phi ^{-}\rangle \right) $ as ''\textit{Visibility}'' of
a superposition $|\phi ^{\pm }\rangle $ is legitimate insofar as the
component states of the corresponding superposition, $|\phi _{1}\rangle $
and $|\phi _{2}\rangle $ may be defined, at least approximately, as ''%
\textit{pointer states}'' or ''\textit{einselected states}'' \cite{Zure03}.
Within the set of the eigenstates characterizing any quantum system the
pointer states are defined as the ones least affected by the external noise
and that are highly resilient to decoherence. In other words, the pointer
states\ are ''quasi classical'' states which realize the minimum flow of
information from (or to) the System to (or from) the Environment. Amongst
the reasonable criteria of classicality, such as the ones based on
''purity'' of the macrostates or on their ''predictability'', discussed by
W. H. Zurek in Reviews of Modern Physics \cite{Zure03}, the
distinguishability criterion adopted in the present work is likely to be
related to the ''\textit{distinguishability sieve''} suggested by
B.W.Schumaker in 1999 and referred to in that paper as a ''private
communication''.

\textbf{The lossy Channel} is modelled in the present analysis by a generic
linear beam-splitter (BS)\ with transmittivity $T$ and reflectivity $R=1-T$
acting on a generic quantum state associated with a single mode beam: Fig.%
\ref{fig:fig1} \cite{Loud,Leon93}. As usually done with photons, the
scattering provided by BS (BS-scattering) is assumed to represents well the
decoherence process, the one that provides the flow of information from the
system to the environment. As it is well known, the BS-scattering is also
generally assumed to model the necessarily limited quantum efficiency of any
realistic photodetector, $QE<1$ \cite{Loud}. Then our present interpretation
may be thought of as to include conceptually the latter effect into an
overall decoherence scheme involving only ideal detectors $(QE=1)$ at the
end of the measurement chain. The calculation of the output density matrix
consists of the $T-$dependent BS-scattering transformation on the input
state, and of the evaluation of the partial trace (R-trace) of the emerging
field on the reflected mode, i.e. the loss variables. The aim of the paper
is to study the evolution of two Macro-states $|\phi _{1}\rangle $ and $%
|\phi _{2}\rangle \ $ and of their superpositions $|\phi _{\pm }\rangle $ by
the size of the corresponding $D\left( |\phi _{1}\rangle ,|\phi _{2}\rangle
\right) $, $D\left( |\phi ^{+}\rangle ,|\phi ^{-}\rangle \right) $ as a
function of the parameters $R$ (or $T)$ of the lossy channel. We start by
analyzing the well known MQS generated by the Glauber's coherent states \cite%
{Brun92,Raim01}. Then we shall consider somewhat extensively the QI-OPA
solution, which is at the focus of the present analysis. At last, the
comparison between the two cases will provide an insightful assessment of
our results.

\section{Quantum Superposition of \ Glauber's Coherent states}

\begin{figure}[t]
\includegraphics[width=0.5\textwidth]{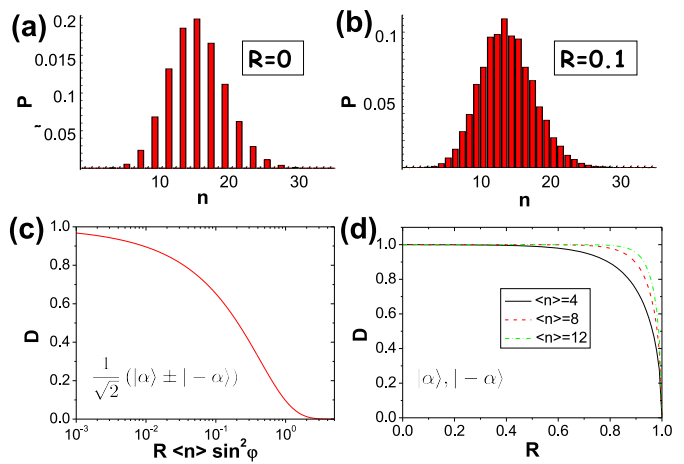}
\caption{(Color online) (a)-(d): Plot of the distribution of the number of photons in the $|%
\protect\phi ^{+}\rangle $ state for $\protect\alpha =4$, corresponding to
an average number of photons $\langle n\rangle =16$, for reflectivities $R=0$
(fig.2-a) and $R=0.1$ (fig.2-b). (c): Plot of the universal curve that
describes the distance between $|\protect\phi ^{+}\rangle $ and $|\protect%
\phi ^{-}\rangle $ after losses as a function of $R\langle n\rangle \sin ^{2}%
\protect\varphi $. This universal curve is valid for the more general case
of the $\vert \protect\phi_{\protect\varphi}^{\pm} \rangle$ states when
plotted as a function of $R \langle n \rangle \sin^{2} \protect\varphi$.
(d): Distance between the coherent $\vert \protect\alpha \rangle$ and $\vert
- \protect\alpha \rangle$ states as a function of the reflectivity $R$ for
different values of $\langle n \rangle$. From lower to the upper curves, the
lines correspond to increasing values of $\langle n \rangle$ as indicated in
the legend.}
\label{fig:fig2}
\end{figure}

The method introduced in the previous Section for studying the resilience to
decoherence of macroscopic states and their quantum superposition is here
applied to the MQS\ of "coherent" Glauber's - states. The properties of the
states have been widely studied in the past \cite{Brun92,Raim01,Schl91} and
represent a crucial test to verify the validity of our present approach. The
MQS made by coherent states $|\phi _{1,2}\rangle =|\pm \alpha \rangle $ is
defined here as: $|\phi ^{\pm }\rangle =\frac{\mathcal{N}}{\sqrt{2}}\left(
|\alpha \rangle \pm |-\alpha \rangle \right) $, where $\mathcal{N}$ is a
normalization quantity \cite{Schl91}. In the specific case of input coherent
states $|\pm \alpha \rangle $, the application of the above loss model leads
after BS - scattering to the output coherent-state density matrix: $\widehat{%
\rho }_{\pm \alpha }^{T}=|\pm \sqrt{T}\alpha \rangle \,\langle \pm \sqrt{T}%
\alpha |$, i.e. the \textquotedblright decoherence\textquotedblright\ due to
scattering doesn't change the structure of the states. The distance between
the two states with opposite phase is easily found \cite{Loud} : $D\left( |%
\sqrt{T}\alpha \rangle ,|-\sqrt{T}\alpha \rangle \right) =\sqrt{%
1-e^{-2T|\alpha |^{2}}}$, a value close to 1 for an average number of
transmitted particles $T|\alpha |^{2}$ where: $0<T|\alpha |^{2}<|\alpha
|^{2} $ Fig.\ref{fig:fig2}-(d). In this regime the coherent states $|\pm
\alpha \rangle $ keep their mutual distinguishability through the lossy
channel and comply with Zurek's definition of \textquotedblright pointer
states\textquotedblright .

Let's now consider the MQS "Visibillity". The R-reduced density matrices
after the BS-scattering have the general form: $\hat{\rho}_{T\pm }=\frac{1}{2%
}\left( |\beta \rangle _{{}}\,_{{}}\langle \beta |+|-\beta \rangle
_{{}}\,_{{}}\langle -\beta | \pm e^{-2R|\alpha |^{2}}\left( |-\beta \rangle
_{{}}\,_{{}}\langle \beta |+|\beta \rangle _{{}}\,_{{}}\langle -\beta
|\right) \right)$ with $|\beta \rangle =|\alpha \sqrt{T}\rangle $. For the
coherent state MQS with no losses ($T=1$), the distribution in the Fock
space exhibits only elements with an even number of photons for $|\phi
^{+}\rangle $ or an odd number of photons for $|\phi ^{-}\rangle $. The
quantum superposition of these states is exclusively attributable to this
very peculiar ''\textit{comb''} structure of the Fock spectrum. This
structure is indeed extremely fragile under the effect of losses since the\
R-trace operation must be carried out in the space of the \textit{%
non-orthogonal }coherent-states. This is shown in Fig.\ref{fig:fig2}(a)-(b)
for increasing values of \ the particle loss. The MQS\ Visibility of these
states can be evaluated analytically in closed form and is found extremely
sensitive to decoherence \cite{nota}:

\begin{equation}
D=\sqrt{1-\sqrt{1-e^{-4R|\alpha |^{2}}}}
\end{equation}

\noindent i.e., $D(x)\simeq e^{-2x}$ being: $x\equiv R<n>=R|\alpha |^{2}\geq
1$, the average number of lost photons. The loss of 1 photon, on the
average, leads to the MQS\ Visibility value: $D=0.096$, and then to the
practical cancellation of any detectable interference effects involving $%
\widehat{\rho }_{\phi ^{\pm }}^{T}$.\ This is fully consistent with the
experimental observations \cite{Brun92,Raim01}. The previous calculations
generalize to the general coherent state MQS: $|\phi _{\varphi} ^{\pm
}\rangle {}$= $\frac{\mathcal{N}_{\varphi }^{\pm }}{\sqrt{2}}\left( |\alpha
e^{\imath \varphi }\rangle \pm |\alpha e^{-\imath \varphi }\rangle \right) $
by substituting $|\alpha |^{2}$ with $|\alpha |^{2}\sin ^{2}\varphi$. Hence,
we can summarize the theoretical results for the MQS\ Visibility by tracing
the \textit{unique} function:\ $D(|\phi _{\varphi +}\rangle {},|\phi
_{\varphi -}\rangle {})=D(x)$ with $x=R|\alpha |^{2}\sin ^{2}\varphi $,
shown in Fig. 2-(c). We consider this ''universal''\ function an additional
important property of the ''coherent states'' (not previously discovered, to
our knowledge). Note that the function $D(x)$ approaches its minimum value\
with zero \textit{slope}: $Sl=\lim_{R\rightarrow 1}\left| dD(x)/dx\right|
=0. $

As noted by Gunnar Bjork \cite{Bjor08}, a similar behavior is obtained by
applying the present decoherence model to the entangled two-mode number
states, i.e. the states $|\phi _{N1}\rangle = $ $|N0\rangle $ and $|\phi
_{N2}\rangle = $ $|0N\rangle $ and to their superpositions, called NOON
states: $\ |\phi _{N\pm }\rangle =$ $\frac{\mathcal{N}^{\pm }}{\sqrt{2}}$ $%
(|N0\rangle $ $\pm $ $|0N\rangle )$, for large N \cite{NOON}. After
R-tracing, the distance corresponding to the Fock-states $|N0\rangle$ and $%
|0N\rangle$ above is found to scale as $\mathcal{D(}\phi _{N1},\phi _{N2})=%
\sqrt{1-R^{N}}$, while the distance corresponding to their superpositions is
found: $\mathcal{D(}\phi _{N\pm })=\sqrt{(1-R)^{N}}$. Then it turns out
that, while $\mathcal{D(}\phi _{N1},\phi _{N2})\simeq 1$ for $\langle
n\rangle \ll N$, the total number of particles, $\mathcal{D(}\phi _{N\pm })$
drops to zero as soon as $R\langle n\rangle \simeq 1$, with the cancellation
of the visibility after loss of a single photon on the average. This
behaviour allows to identify $\phi _{N1},\phi _{N2}$ with the "pointer
states" of the system, which show a slow decoherence rate and can in
principle be discriminated bu a suitable measurement even in a high lossy
transmission channel similarly to the $\vert \alpha \rangle$ states. On the
contrary, the quantum superpositions $\phi _{N\pm }$ are fragile under
decoherence as the $(\vert \alpha \rangle \pm \vert -\alpha \rangle)$ states
are.


\begin{widetext}

\begin{figure}[hb]
\centering
\includegraphics[width=0.7\textwidth]{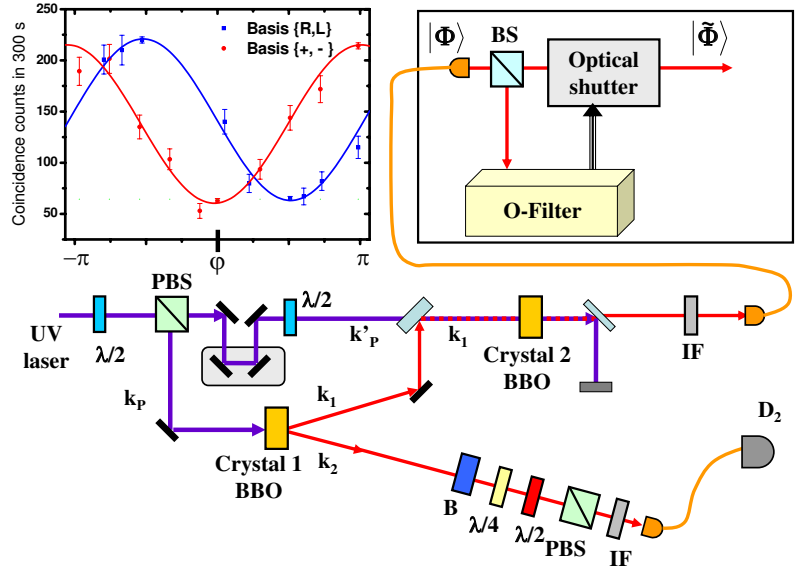}
\caption{(Color online) Scheme of the experimental setup for the observation of entanglement 
between a microscopic and a macroscopic system. The
main UV laser beam provides the OPA excitation field beam at $\lambda = 397.5 nm$. A type II
BBO crystal (crystal 1: C1) generates pair of photons with
$\lambda = 795 nm$. In virtue of the EPR non-local correlations established between the modes $\mathbf{k}_{1}$ 
and $\mathbf{k}_{2}$, the preparation of a single-photon on mode $\mathbf{k}_{1}$ with 
polarization state $\vec{\pi}_{\varphi}$ is conditionally determined
by detecting a single-photon  after proper 
polarization analysis on the mode $\mathbf{k}_{2}$ (polarizing beamsplitter (PBS), $\lambda/2$ and 
$\lambda/4$ waveplates, Soleil-Babinet compensator (B), interferential filter (IF)). 
The photon belonging to $\mathbf{k}_{1}$, together with the pump laser beam $\mathbf{k}_{p}'$, 
is fed into an high gain optical parametric amplifier
consisting of a NL crystal 2 (C2), cut for
collinear type-II phase matching. The fields are coupled to single-mode fibers. 
For more details refer to \cite{DeMa08}.
\textbf{Left inset}: Experimental results 
of the interference fringe pattern between the microscopic $\mathbf{k}_{2}$ 
and the macroscopic $\mathbf{k}_{1}$ fields \cite{DeMa08}. Square data corresponds
to the fringe pattern in the $\{R,L\}$ basis, circular data in the $\{+,-\}$ basis.
\textbf{Right Inset}: O-Filtering process obtained by an
"idle" measurement apparatus. A portion of the wave-function ($\simeq 10\%$) is measured 
and analyzed by an O-Filter driving a fast e-optical shutter.}
\label{fig:fig3}
\end{figure}

\end{widetext}

In conclusion, note that the very high resilience to decoherence shown by
the ''pointer states'' just considered parallels a very high sensitivity to
decoherence of the quantum superpositions of the same states. As we shall
see, this important property is at variance with the behavior of the
''pointer states'' realized by Quantum Cloning.

\section{Quantum superposition by Optimal phase-covariant Quantum Cloning.}

In this Section we address the central point of the paper. As said, recent
experimental results \cite{DeMa08} showed that quantum properties, such as
quantum entanglement, can be still observed in macroscopic system of $%
\approx 10^{4}$ particles even after the transmission over a lossy channel
and the detection by a measurement apparatus with non-unitary detection
efficiency. In facts, according to the experimental evidence and in
agreement with the theory reported in this Section, an exceedingly high
resilience to decoherence is common both to the\ \textquotedblright
pointer\textquotedblright\ macrostates generated by the QI-OPA system and,
by the same amount, to all quantum superpositions of the same states \cite%
{DeMa08}. In order to give a theoretical insight on this task, we apply the
method of Sec.II to the amplified single photon qubits by a collinear
QI-OPA, i.e. the simplest \textit{\textquotedblright optimal phase-covariant
quantum cloning machine\textquotedblright }. We first derive the expression
of the density matrix of these multi-photon states after propagation over a
lossy channel, for several input polarization states of the injected qubit.
\ The evolution of the photon-number distributions leads to a
first insightful picture of the effects of decoherence and introduces the
more quantitative results given by the Bures distance. The latter is then
evaluated by performing a numerical calculation of the quantum fidelity
between the exact density matrices. Furthermore, the same procedure is
applied to the recently discovered O-Filter device \cite{DeMa08} showing
that this device can substantially reduce the decoherence effects on the MQS
visibility, at the cost of discarding a part of the data.

\begin{widetext}

\begin{figure}[ht]
\includegraphics[width=0.7\textwidth]{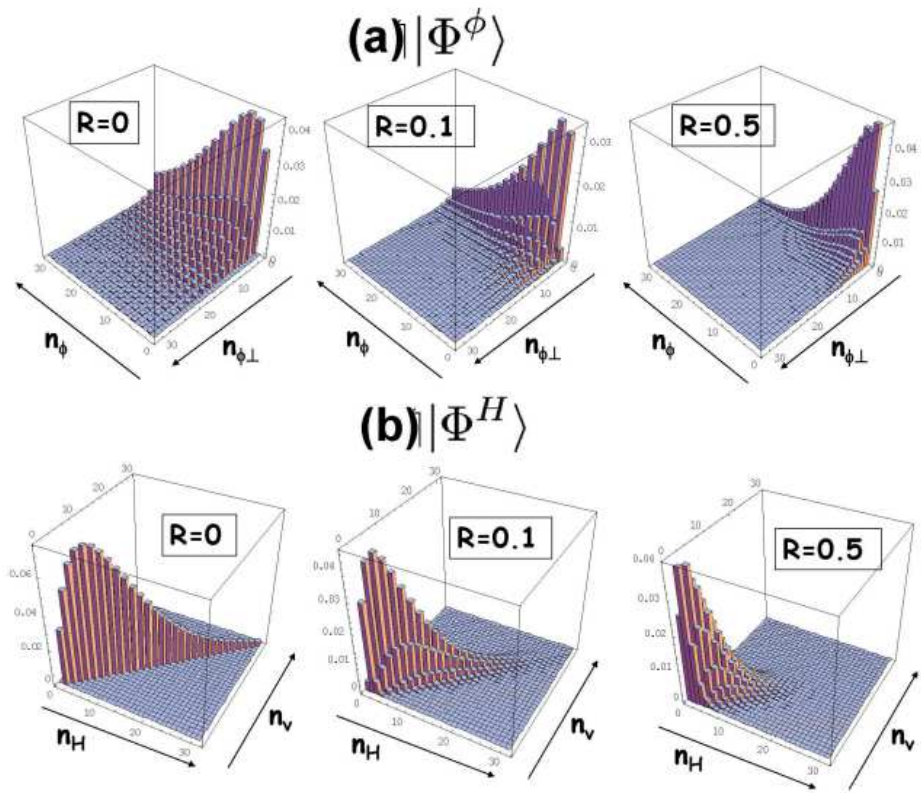}
\caption{(Color online) (a) Probability distribution in the Fock space $(n_{\protect\phi
},n_{\protect\phi _{\bot }})$ for the amplified $|\Phi ^{\protect\phi
}\rangle $ state of a generic equatorial qubit for different values of the
transmittivity. (b) Probability distribution in the Fock space $%
(n_{H},n_{V}) $ for the amplified $|\Phi ^{H}\rangle $ state for different
values of the transmittivity. All distributions refer to a gain value of $%
g=1.5$, corresponding to an average number of photons $\langle n\rangle
\approx 19$.}
\label{fig:fig4}
\end{figure}

\end{widetext}

Let us begin by describing the system under investigation. The QI-OPA\
device is represented in Fig.\ref{fig:fig3} together with the interference
fringes of the Macrostate quantum superpositions obtained in a recent
experiment (Left Inset) \cite{DeMa98,DeMa08}. Remind that the \textit{%
phase-covariant} process clones identically and ''\textit{optimally}'', i.e.
with the minimal ''\textit{squeezed - vacuum noise}'' allowed by the \textit{%
no-cloning} theorem and implied by parametric amplification, all input
qubits belonging on the Poincar\'{e} sphere to the ''equatorial''\ plane
orthogonal to the polarization basis $\left\{ \vec{\pi}_{H},\vec{\pi}%
_{V}\right\} $. Here H and V refer to horizontal and vertical optical
polarizations \cite{Scia05,Scia07}. \ The interaction Hamiltonian is: $%
\widehat{\mathcal{H}}_{coll}$=$\imath \hbar \chi \widehat{a}_{H}^{\dag }%
\widehat{a}_{V}^{\dag }+\mathrm{h.c.}$ when expressed in the basis $\left\{ 
\vec{\pi}_{H},\vec{\pi}_{V}\right\} $, or $\widehat{\mathcal{H}}_{coll}$=$%
\frac{\imath \hbar \chi }{2}e^{-\imath \phi }\left( \widehat{a}_{\phi
}^{\dag \,2}-e^{\imath 2\phi }\widehat{a}_{\phi _{\bot }}^{\dag \,2}\right) $
when expressed in any equatiorial basis $\left\{ \vec{\pi}_{\phi },\vec{\pi}%
_{\phi \perp }\right\} $, where: $\vec{\pi}_{\phi }=2^{-\frac{1}{2}}\left( 
\vec{\pi}_{H}+e^{\imath \phi }\vec{\pi}_{V}\right) $. Two relevant
''equatorial'' bases $\left\{ \vec{\pi}_{+},\vec{\pi}_{-}\right\} $ and $%
\left\{ \vec{\pi}_{R},\vec{\pi}_{L}\right\} $ correspond respectively to the
phase sets $\phi =\left\{ 0,\pi \right\} $ and $\phi =\left\{ \pi /2,3\pi
/2\right\} $. By direct calculation, obtained applying the unitary cloning
operator $\hat{U}=e^{-\frac{\imath \mathcal{H}_{int}t}{\hbar }}$, the output
states for an injected qubit $\pi =\left\{ H,V\right\} $ is found to have
the Fock-state expansion \cite{DeMa08}:

\begin{equation}  \label{eq:linear_unperturbed}
|\Phi ^{\pi }\rangle =\hat{U}|\pi \rangle =\frac{1}{C^{2}}\sum_{i=0}^{\infty
}\Gamma ^{i}\sqrt{i+1}\,|(i+1)\pi ,i\pi _{\bot }\rangle
\end{equation}
where the ket $\vert n \pi, m \pi_{\bot} \rangle$ represents the number
state with $n$ photons with $\pi$ polarization and $m$ photons with $%
\pi_{\bot}$ polarization. With the same procedure, for an injected
equatorial qubit the \textit{optimally} amplified state is: 
\begin{equation}
|\Phi ^{\phi }\rangle =\hat{U}|\phi \rangle =\sum_{i,j=0}^{\infty }\gamma
_{ij}|(2i+1)\phi ,(2j)\phi _{\bot }\rangle  \label{eq:Phi_equat}
\end{equation}

\noindent where $\gamma _{ij}=\frac{1}{C^{2}}\left( e^{-\imath \varphi }%
\frac{\Gamma }{2}\right) ^{i}\left( -e^{\imath \varphi }\frac{\Gamma }{2}%
\right) ^{j}\frac{\sqrt{(2i+1)!}\,\sqrt{(2j)!}}{i!j!}$. In these expressions 
$C=\cosh g$ and $\Gamma =\tanh g$, where g is the non linear gain of the
amplifier. \ Consider the ''equatorial'' macrostates $|\Phi ^{+}\rangle $
and $|\Phi ^{-}\rangle $ corresponding respectively to $\phi =0$ and $\phi
=\pi $, and let's assume them provisionally as ''pointer macrostates'' \cite%
{Zure03}. The general expression of any macroqubit lying on the equatorial
plane may be taken as a MQS of $|\Phi ^{+}\rangle $ and $|\Phi ^{-}\rangle $%
: $|\Phi ^{\phi }\rangle =e^{-i\phi /2}[\cos (\phi /2)|\Phi ^{+}\rangle $ + $%
i\sin (\phi /2)|\Phi ^{-}\rangle ]$. Assume now, for the sake of
definiteness, the two independent MQS's identified by the new ''equatorial''
basis $\left\{ |\Phi ^{R}\rangle ,|\Phi ^{L}\rangle \right\} $: $|\Phi
^{R}\rangle =\frac{\mathcal{N}_{\pm }}{\sqrt{2}}\left( |\Phi ^{+}\rangle
+i|\Phi ^{-}\rangle \right) $ and $|\Phi ^{L}\rangle =\frac{\mathcal{N}_{\pm
}}{\sqrt{2}}\left( |\Phi ^{+}\rangle -i|\Phi ^{-}\rangle \right) $ \cite%
{DeMa05}. Indeed, owing to linearity and to the phase-covariance of the
cloning process, each basis set of macrostates lying on the equatorial plane
is a quantum superposition of any other macrostate set lying on that same
plane. Therefore the \textit{distinguishability of }$\left\{ |\Phi
^{+}\rangle ,|\Phi ^{-}\rangle \right\} $ expressed by the distance $D(|\Phi
^{+}\rangle ,|\Phi ^{-}\rangle )$\ coincides with\ the MQS\ \textit{%
Visibility} of any superpositions $|\Phi ^{\phi }\rangle $, as for instance
of $|\Phi ^{R}\rangle $ or $|\Phi ^{L}\rangle $:

\begin{equation}
D(|\Phi ^{R}\rangle ,|\Phi ^{L}\rangle )=D(|\Phi ^{+}\rangle ,|\Phi
^{-}\rangle )
\end{equation}

In conclusion, the peculiar phase-covariant symmetry of the cloning process
allows to identify the equatorial plane of the Poincar\`{e} sphere of the
macroqubits as a preferred Hilbert subspace in which the assumed ''pointer''
macrostates as well as any MQS contributed by them are affected by the same
decoherence process and are then granted by an identical \textit{%
einselection }property \cite{Zure03}. As already noted, all these important
properties of the QI-OPA\ system are at variance with the general case
considered by Einstein in the quoted letter to Born and with the particular
case with coherent or NOON\ states just considered.

In order to assess the einselected status of the macrostates at hand and the
related MQS's, let's now consider the decoherence induced by BS-scattering
and the corresponding Fock space spectra. The first step is the calculation
of the explicit form of the density matrix after transmission over the lossy
channel for the equatorial Macro-qubits. 
We report here their explicit form: 
\begin{widetext}
\begin{equation}
\label{eq:density_equatorial_loss}
\begin{aligned}
\left( \hat{\rho}^{\phi}_{T} \right)_{ijkq} &= \sum_{m,n=0}^{\infty} \frac{1}{C^{4}}
\left( \frac{\Gamma}{2} \right)^{\frac{i+j+k+q}{2}+m+n-1} 
\left( -1 \right)^{\frac{j+q}{2}+n} \left( e^{\imath \varphi} \right)^{\frac{j+k-i-q}{2}} 
\frac{\sqrt{(i+m)!(j+n)!(k+m)!(q+n)!}}{\left(\frac{i+m-1}{2}\right)! \left(\frac{j+n}{2}\right)!
\left(\frac{k+m-1}{2}\right)! \left(\frac{q+n}{2}\right)!} \\
&\left( \sqrt{T} \right)^{i+j+k+q} 
\left( R \right)^{m+n}
\left[ \begin{pmatrix} i+m \\ m \end{pmatrix} \begin{pmatrix} j+n \\ n \end{pmatrix}
\begin{pmatrix} k+m \\ m \end{pmatrix} \begin{pmatrix} q+n \\ n \end{pmatrix} \right]^{\frac{1}{2}} 
f(i,k,m) h(j,q,n)
\end{aligned}
\end{equation}
\end{widetext}
corresponding to the $|i\phi ,j\phi _{\bot }\rangle \,\langle
k\phi ,j\phi _{\bot }|$ matrix element, where $f(i,k,m)$ and $h(j,q,n)$ express the 
constraints over the parity of the indexes $\left\{ i,j,k,q \right\}$ according to: 
\begin{equation}
f(i,k,m)=\left\{ 
\begin{array}{ll}
1 & \mathrm{if}\,i\,mod(2)=k\,mod(2)\neq m\,mod(2) \\ 
0 & \mathrm{otherwise}%
\end{array}%
\right.
\end{equation}%
\begin{equation}
h(j,q,n)=\left\{ 
\begin{array}{ll}
1 & \mathrm{if}\,j\,mod(2)=q\,Mod(2)=n\,mod(2) \\ 
0 & \mathrm{otherwise}%
\end{array}%
\right.
\end{equation}%
We stress that both these constraints and the sums $\sum_{m}\sum_{n}$ derive
from the interaction of the initial wave function with the beam-splitter $%
\hat{U}_{BS}$ \cite{Loud} and the subsequent R-tracing, i.e. the partial
trace over the reflected mode of $\hat{U}_{BS}|\Phi ^{\phi }\rangle $. These
sums however can be further rearranged by use of the Hyper-geometric
functions $_{2}F_{1}(\alpha ,\beta ;\gamma ;z)$ \cite{Slat66}.

With an analogous procedure, the $\left\{ \pi =H,V\right\} $ states
after the R-tracing are described by the following density matrix: 
\begin{equation}
\begin{aligned} &\hat{\rho}^{H}_{T} = \sum_{i=1}^{\infty} \sum_{j=0}^{i-1}
\sum_{k=0}^{\infty} \left( \sum_{p=0}^{\infty} \overline{\gamma}_{ijk;p}
\right) \vert i\pi ,j\pi_{\bot} \rangle \, \langle k\pi , (k+j-i)\pi_{\bot}
\vert +\\ &+ \sum_{i=0}^{\infty} \sum_{j=i}^{\infty} \sum_{k=0}^{\infty}
\left( \sum_{p=j+1-i}^{\infty} \overline{\gamma}_{ijk;p} \right) \vert i\pi
,j\pi_{\bot} \rangle \, \langle k\pi , (k+j-i)\pi_{\bot} \vert \end{aligned}
\end{equation}%
where the coefficients $\overline{\gamma }_{ijk;p}$ are: 
\begin{equation}
\begin{aligned} \overline{\gamma}_{ijk;p} &= \frac{\Gamma^{2p+i+k-2}}{C^{4}}
\sqrt{p+i} \sqrt{p+k} \, T^{k+j} R^{2p+i-1-j} \\ &\left[\begin{pmatrix} p+i
\\ i \end{pmatrix} \begin{pmatrix} p+i-1 \\ j \end{pmatrix} \begin{pmatrix}
p+k \\ k \end{pmatrix} \begin{pmatrix} p+k-1 \\ k+j-1
\end{pmatrix}\right]^{\frac{1}{2}} \end{aligned}
\end{equation}

Fig.\ref{fig:fig4} reports the distribution in the Fock space $P(n_{\phi
,}n_{\phi \perp })$ corresponding to different macrostates $|\Phi ^{\phi
}\rangle $ for different values of the reflectivity $R.$ For the unperturbed
states (R=0), as shown by Fig.\ref{fig:fig4}-(a), each \textquotedblright
equatorial\textquotedblright\ macrostate $|\Phi ^{\phi }\rangle $\ ,
evaluated by Eq.(\ref{eq:Phi_equat}) exhibits a typical \textit{comb}
structure, i.e. the spectrum of Fock space contains only\ nonvanishing terms
with a specific parity, in particular odd photon numbers for $\vec{\pi}%
_{\phi }$ polarization and even photon numbers for its orthogonal $\vec{\pi}%
_{\phi }{}_{\bot }$. On the other hand, the amplified $\left\{ \left\vert
\Phi ^{H,V}\right\rangle \right\} $ states, that are not equatorial states,
are characterized by a diagonal distribution. When losses are inserted (Fig.%
\ref{fig:fig4}), these peculiar properties are progressively cancelled, as
for the Glauber's - states MQS considered in Section III. However, the
distributions corresponding to initially orthogonal macroqubits remain
distinguishable even after losses only for the equatorial $|\Phi ^{\phi
}\rangle $ states (Fig.\ref{fig:fig4}-(a)), since most of the events are
localized in different Fock-space zones, i.e. where the number of \
amplified photons bearing the same polarization $\overrightarrow{\pi }$ of
the injected qubit is substantially higher than the ones which are in the
orthogonal $\overrightarrow{\pi}_{\pi}$. This indeed corresponds to the optimal
quantum cloning feature of the QIOPA, that in the high gain regime survives
in lossy schemes. This latter feature is absent in the $|\Phi ^{H,V}\rangle $
macroqubits since the $\left\{ \vec{\pi}_{H},\vec{\pi}_{V}\right\} $ basis,
is not "equatorial" and then it doesn't correspond to\ an "optimal"\ cloning
by a collinear QI-OPA.

\begin{figure}[t]
\centering
\includegraphics[scale=.5]{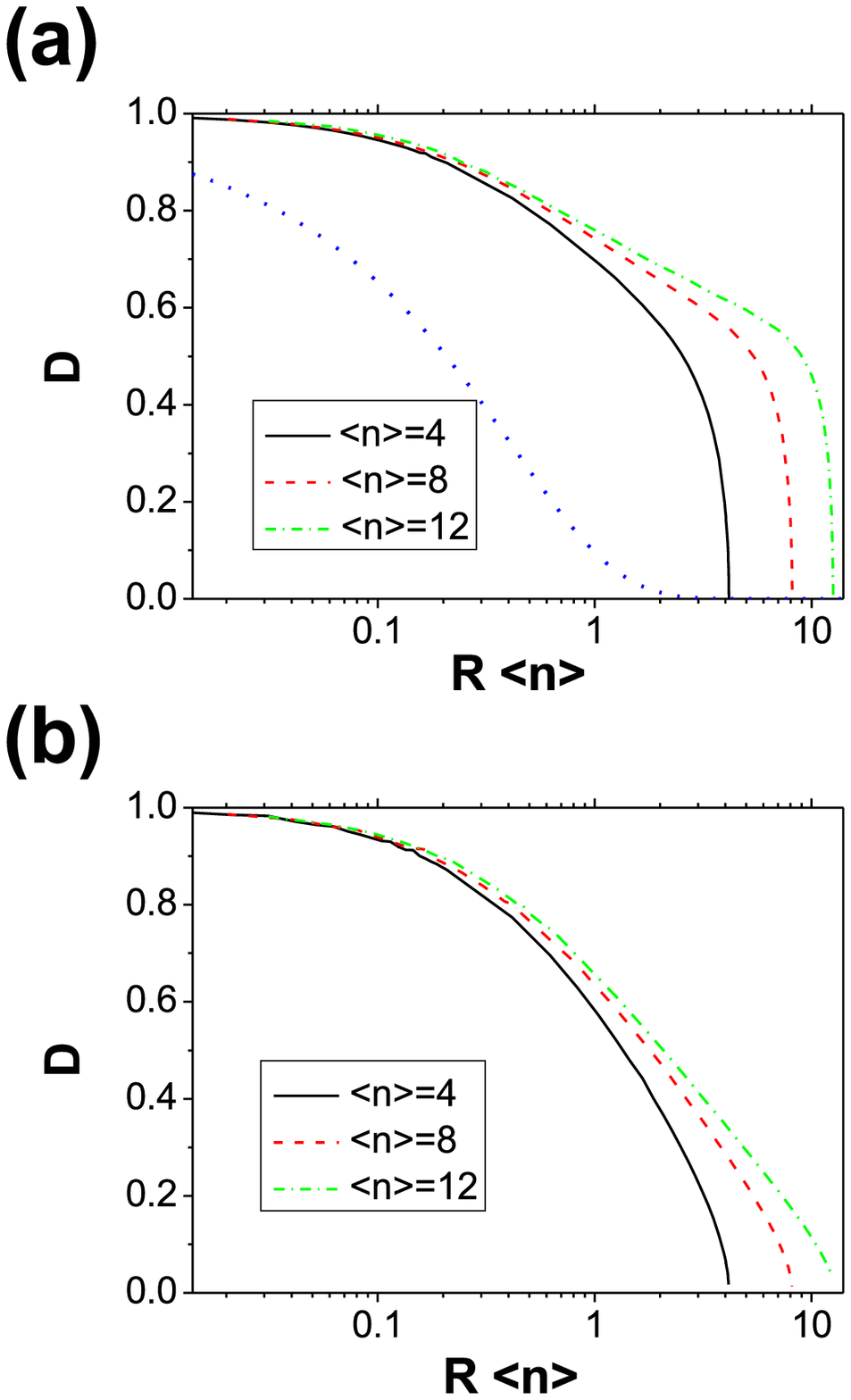}
\caption{(Color online) (\textbf{a}) Numerical evaluation of the distance $D(x)$ between
two orthogonal equatorial macro-qubits $|\Phi^{\protect\phi,\protect\phi%
_{\bot}}\rangle$ as function of the average lost particle $x=R<n>$, plotted
in a logarirhmic scale. Black straight line refers to $g=0.8$ and hence to $\langle n
\rangle \approx 4$, red dashed middle line to $g=1.1$ and $\langle n \rangle \approx 8$,
green dash-dotted upper line to $g=1.3$ and $\langle n \rangle \approx 12$. The blue dotted
lower line corresponds to the function $D(x)$ for coherent state MQS, universal for any
value of $\langle n \rangle$. (\textbf{b}) Numerical evaluation of the Bures
distance between $\vert \Phi^{H} \rangle$ and $\vert \Phi^{V} \rangle$ for
the same values of the gain of (a). Black straight lower line refers to $g=0.8$ and hence to $\langle n
\rangle \approx 4$, red dashed middle line to $g=1.1$ and $\langle n \rangle \approx 8$,
green dash-dotted upper line to $g=1.3$ and $\langle n \rangle \approx 12$. 
We note the faster decrease of the distance with respect to the equatorial case.}
\label{fig:fig5}
\end{figure}

\begin{figure}[h]
\centering
\includegraphics[width=0.4\textwidth]{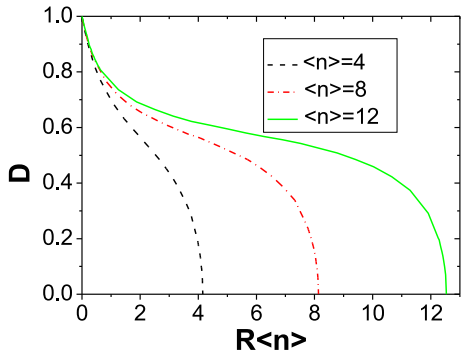}
\caption{(Color online) Numerical evaluation of the Bures distance D(x) between two
orthogonal equatorial macro-qubits $|\Phi^{\protect\phi,\protect\phi%
_{\bot}}\rangle$ as function of the average lost particle $x=R<n>$, plotted
in linear scale. Black dashed lower line refers to $g=0.8$ and hence to $\langle n
\rangle \approx 4$, red dash-dotted middle line to $g=1.1$ and $\langle n \rangle \approx 8$,
green straight upper line to $g=1.3$ and $\langle n \rangle \approx 12$. As the number of
particles generated increases, two different decoherence regimes can be
discriminated, i.e. in the low and high loss range.}
\label{fig:fig6}
\end{figure}

The \textit{visibility} $D(x)$ of the MQS belonging to the ''equatorial''
subspace $\left\{ |\Phi ^{\phi }\rangle ,|\Phi ^{\phi \bot }\rangle \right\} 
$ has been evaluated numerically as function of the average\textit{\ lost}
photons: $x\equiv R<n>$. More specifically, we calculated the quantum
fidelity between these state by a numerical program which re-organized the
density matrix of eqq.(\ref{eq:density_equatorial_loss}) in a linear
algebraic matrix form. The fidelity, and hence the Bures distance, was then
estimated by using algebraic numerical routines. The results for different
values of the gain are reported in Fig.\ref{fig:fig5}-(\textbf{a}). Note
that for small values of $x $ the decay of $D(x)$ is far slower than for the
coherent state case shown in Fig.\ref{fig:fig2}-(\textbf{e}) and reproduced
again in Fig.\ref{fig:fig5} for comparison (dotted line). A direct
comparison with the previous case shows that the resilience of QIOPA
amplified states increases with a higher number $N$, the total output number
of particles in the primary beam. Furthermore, after a common inflexion
point at $D\sim 0.5$ the\ function $D(x)$ drops to zero for $R=1$ and then
for: $<n>\sim N$. Very important, for large $<n>$, i.e. $R\rightarrow 1$ the 
\textit{slope} of the functions $D(x)$ increase fast towards a very large
value: $R\rightarrow 1$: $Sl=\lim_{R\rightarrow 1}\left| dD(x)/dx\right|
\approx \infty $. The latter property can be demonstrated considering the
high lossy regime for $R \approx 1$, and keeping only the first order terms
in $T$ of the density matrix, obtaining: $\lim_{T \rightarrow 0} \frac{%
\partial D(\hat{\rho}_{T}^{\phi}, \hat{\rho}_{T}^{\phi_{\bot}})}{\partial T} = 1 + 4
C^{2} + 2 C^{2} \Gamma (1 + 2 \Gamma^{2})^{\frac{1}{2}}\overset{g\rightarrow
\infty}{\rightarrow}\infty$

All this means that the MQS\ Visibility can be large even if the average
number $x$ of lost particles is close to the total number $N$, i.e. for $%
R\sim 1$. As seen, this behavior is opposite to the case of \textit{coherent
states} where the function $D(x)$ approaches zero value with \textit{zero}
slope: Fig.2-(e). We believe that this lucky and quite unexpected behavior
is at the core of the high resilience to decoherence of our QI-OPA MQS\
solution. Note that this behavior was responsible for the well resolved
interference pattern with visiblity: $V\approx 20\%$ obtained in absence of
O-Filter (OF) by \cite{Naga07}. As a trivial remark, note that in all
Figures of the prseent work the function $D(x)$ necessarily drops to zero
for $<n>\simeq N$, the total number of particles associated to the
macrostates.

To gain insight on this feature we report in Fig.\ref{fig:fig6} the \emph{%
visibility} between the $|\Phi ^{\pm }\rangle $ states as a function of the
average lost photons plotted in a \textit{linear scale}. As the number of
photons generated by the amplification increases, the curves allows to
identify two different regimes in the decoherence process. As said, at low $%
R $, the rapid decay is due to the cancellation of the \textit{comb}
structure in the photon number distribution. On the contrary, when $R$ is
progressively increased, the initial exponential decay is interrupted, a
kind of \textit{plateau} appears and a more resilient structure is found.
This resilient portion of the spectrum is attributable to the unbalancement
in the photon number distributions in the Fock space $P(n_{\Phi ,}n_{\Phi
\perp })$ for the ''equatorial''\ macrostate $|\Phi ^{\phi }\rangle $ due to
the polarization encoding of the \textit{seed} microqubits. This kind of
encoding is missing in the coherent-state MQS\ case since there the fragile
MQS interference is only related to the existence of the \textit{comb}
structure of the Fock spectra and then quickly disappears with it, as said.
In other words, two initially orthogonal polarization states maintain their
distributions unbalanced in different zones of the two-dimensional Fock
space even in regimes of large losses. An interesting case is presented by
Fig. \ref{fig:fig5}-(b) that shows the rapid decay of the coherence of $%
D(|\Phi ^{H}\rangle ,|\Phi ^{V}\rangle )$ under BS-scattering of the
macrostates $|\Phi ^{H}\rangle ,|\Phi ^{V}\rangle $ which \textit{do not
belong} to the ''equatorial plane'' of the Poincar\'{e} sphere. A close
comparison with Fig.\ref{fig:fig5}-(a) emphasizes the role of the privileged
''noise reduction'' Hilbert subspace with respect to decoherence of the
macrostates.

\textbf{Orthogonality Filter (O-Filter)}. The demonstration of
microscopic-macroscopic entanglement by adopting the O-Filter was reported
in \cite{DeMa08}. The POVM like technique \cite{Pere95} implied by this
device locally selects the events for which the difference between the
photon numbers associated with two orthogonal polarizations $|m-n|>k$, i.e.
larger than an adjustable threshold, $k$ \cite{Naga07}. By this method a
sharper discrimination between the output states $|\Phi ^{\phi }\rangle $
e $|\Phi ^{\phi _{\bot }}\rangle $ can be achieved. The action of the
O-Filter can be formalized through the measurement operator $\hat{P}%
_{OF}=\sum_{m,n}|m\phi ,n\phi_{\bot} \rangle \,\langle m \phi,n \phi_{\bot}|$, where the sum over $m,n$
extends over the terms for which the above inequality holds. The O-Filter
can be implemented experimentally either by a post-selected configuration in
the measurement apparatus directly coupled to the output of the QI-OPA
device or by the optical scheme represented in the Right Inset of Fig.\ref%
{fig:fig3}. There a small portion ($\simeq $10\%) of the photon flux
associated with the macrostate $|\Phi ^{\phi }\rangle $ realized at the
output of the QI-OPA is analyzed by an ''idle'' measurement apparatus
connected to an O-Filter that activates a polarization preserving,
high-voltage, fast electro-optical shutter. At last, the O-Filtered
macrostate $|\widetilde{\Phi }^{\phi }\rangle $, corresponding to $|\Phi
^{\phi }\rangle $ emerges at the output of the apparatus \cite{Spag08}. 

\begin{figure}[h]
\centering
\includegraphics[width=0.4\textwidth]{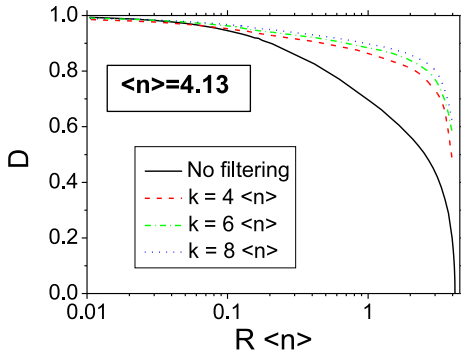}
\caption{(Color online) Numerical evaluation of the Bures distance between two orthogonal
equatorial O-filtered macro-qubits for different values of the threshold $k$
($g=0.8$). Black straight lower line corresponds to the case in which the O-Filter is not
applied. For the remaining curves, from the bottom to the top, red dashed line corresponds to a threshold 
$k = \langle 4 \rangle$, green dash-dotted line to $k = \langle 6 \rangle$ and blue dotted
line to $k = \langle 8 \rangle$.}
\label{fig:fig7}
\end{figure}

We now analyze theoretically the action of the O-Filter device evaluating the
Bures distance between the filtered states $\frac{\hat{P}_{OF} \hat{\rho}_{T}^{\phi} 
\hat{P}_{OF}^{\dag}}{\mathrm{Tr}(\hat{P}_{OF} \hat{\rho}_{T}^{\phi} 
\hat{P}_{OF}^{\dag})}$ and $\frac{\hat{P}_{OF} \hat{\rho}_{T}^{\phi_{\bot}} 
\hat{P}_{OF}^{\dag}}{\mathrm{Tr}(\hat{P}_{OF} \hat{\rho}_{T}^{\phi_{\bot}} 
\hat{P}_{OF}^{\dag})}$. We then applied the numerical methods previously described 
to calculate the Bures distance $D(x)$ as a function of the threshold $k$.
In Fig.\ref{fig:fig7} the results of a numerical analysis carried out for $%
g=0.8 $ and different values of $k$\ are reported. 

Note the increase of the value of $D(x)$, i.e. of the MQS\ Visibility, by
increasing $k$ and, again: $Sl=\lim_{R\rightarrow 1}\left| dD(x)/dx\right|
\approx \infty $. \ Of course, in the spirit of any POVM\ measurement, the
high interference visibility is here achieved at the cost of a lower success
probability \cite{Gisi96}. \ The general, most important feature shown by
all these Figures is that both the ''\textit{Distinguishability}'' and the ''%
\textit{Visibility}'' of all ''equatorial'' macrostates $|\Phi ^{\phi
}\rangle ,|\Phi ^{\phi \bot }\rangle $ as well as of all their
''equatorial'' quantum superpositions can be kept close to the maximum value
in spite of the increasing effect of decoherence due to increasing values of
the quantity: $R<n>$. On the basis of all these results we may then conclude
that all the ''equatorial'' macrostates and superpositions generated by the
QI-OPA\ may be safely defined as classically stable, \textit{einselected} ''%
\textit{pointer states}'' \cite{Zure03}. The validity of this
statement has been demonstrated in the Laboratory, as shown for instance by
the high visibility of the two experimental interference patterns
corresponding to two different measurement orthogonal bases, and appearing
in the Left Inset of Fig.\ref{fig:fig3} \cite{DeMa08}.

\section{Interpretation and conclusions}

The efficiency of the transfer of classical or quantum information in the
interactive dynamics involving the paradigmatic quantum - statistical
combination (\textit{System + Environment}) is at the focus of the present
investigation. \ In order to gain insight into the general picture and to
support the congruence of our final conclusions we find useful to relate
here the various aspects of the cloning process provided by QI-OPA with the
current, most sensible MQS\ physical models, in particular with Zurek's one 
\cite{Zure03}.

1) The ''\textit{System}'' in our scheme is represented by the assembly of $%
N $ photon particles associated with any macrostate $|\Phi ^{\phi }\rangle $
generated by the optical parametric method based on phase-covariant quantum
cloning.

2) The flow of (classical) ''noise information'' directed from the ''\textit{%
Environment}'' towards the \textit{System} is provided in our case by the
unavoidable squeezed-vacuum noise affecting the building up of the
macrostate $|\Phi ^{\phi }\rangle $ within the process of parametric
amplification. \ As already stressed, the ''\textit{optimality}'' of the
phase - covariant quantum cloning adopted in our experiments implies, and
literarly means, that the flow of classical noise is the \textit{minimum}
allowed by the principles of quantum mechanics, i.e. by the ''\textit{%
no-cloning theorem}'' \cite{Scia05,Scia07}.

3) The flow of quantum information directed from the System towards the
Environment is provided by the controlled ''decoherence in action'' provided
by the artificial BS-scattering process adopted for our analysis as well as
by the losses taking place in any realistic photo-detector. We have seen
that by the use of the Orthogonality Filter, or even in the absence of it,
the interference phase-distrupting effects caused by the adopted artificial
decoherence can be efficiently tamed and even cancelled to a great extent
for the ''equatorial'' macrostates and for their quantum superpositions.

4) The selected quantum cloning method, realized by the QI-OPA device,
allows to define a privileged ''\textit{minimum noise - minimum decoherence}%
'' Hilbert subspace of the quantum macrostates that, according to our
decoherence model, exhibit simultaneously the maximum allowed \textit{%
Distinguishability} and \textit{Visibility}. According to the theory of
quantum cloning a smaller size of the privileged Hilbert subspace, here the
equatorial plane of the Poincar\`{e} sphere, corresponds to a higher
''cloning fidelity'' and then to a smaller flux of squeezed - vacuum ''noise
information'' from the environment to the system.

5) The last paragraph of the sentence written by Einstein on the back of his
greetings card to Born appears to conflict with the behavior of the system
generated by our cloning apparatus. \ In facts, in our case quantum
mechanics cannot be taken as ''incompatible'' with the ''classical''
localization of any macro-state $|\Phi ^{\phi }\rangle $ $=\hat{U}|\phi
\rangle $ more than it may be for the ''quantum'' localization of the
corresponding micro-state $|\phi \rangle $. For, in our case a unitary
quantum-cloning transformation $\hat{U}$ connects, we would say ''chains''
albeit in a noisy manner, all physical properties belonging to the
micro-world to the corresponding ones belonging to the macrosopic
''classical'' world. Any lack of perceiving or rationally accepting this
close correspondence, for instance in connection with the realization or
detection of the ''Schr\"{o}dinger Cat'', must be only attributable to the
intrinsic limitations of our perceiving senses, of our observational methods
or of our measurement apparata. In other words, at least in our case, the
two worlds, the Macro and the Micro are deterministically mirrored one into
the other where the ''mirror'', albeit somewhat blurred, is provided by
quantum mechanics itself. In a forthcoming paper many conceptual and
theoretical aspects of the work presented here will be further analyzed in
the phase-space by investigating in details the Wigner functions of the MQS
generated by phase-covariant quantum cloning in presence of losses. 
All these results belongs to the lessons we learned recently by the 
experiments carried out in our Laboratory.

In summary, the present work was intended to give a somewhat firm conceptual
basis to the unexpected high resilience to decoherence
demonstrated in recent experiments by our QI-OPA generated Macroscopic
Quantum Superposition. We believe that this novel MQS system may play
further relevant roles in the future investigations on the foundational
structure of Quantum Mechanics. We acknowledge very useful discussions and
correspondence with Chiara Vitelli and Gunnar Bjork. Work supported by PRIN
2005 of MIUR and INNESCO 2006 of CNISM.


\begin{thebibliography}{99}

\bibitem{Eins65} Einstein-Born Briefwechsel 1916-1955, Nymphenburger Verlagshandlung GmbH, Munchen (1969) 

\bibitem{Eins35} A. Einstein, et al., \emph{Phys. Rev.} \textbf{47}, 777 (1935).

\bibitem{Schr35} E. Schroedinger,  \emph{ Naturwissenschaften} \textbf{23}, $807-812$, $823-828$, $844-849$ (1935).

\bibitem{Niel00} M.A. Nielsen and I.L.Chuang, Quantum Information and Quantum Computation (Cambridge University Press, 2000)

\bibitem{Zure} W. Zurek, Physics Today, October 1991, pag.36; \emph{Rev. Mod. Phys.} \textbf{75}, 715 (2003); \emph{Progr. Math. Phys.}  \textbf{48}, 1 (2007)

\bibitem{Dur02} W. Dur, et al., \emph{Phys. Rev. Lett.} \textbf{89}, 210402 (2002)
; W. Dur, et al., \emph{Phys. Rev. Lett.} \textbf{92}, 180403 (2004)

\bibitem{Gori07} T. Gorin, et al., \emph{Phys. Rev. Lett.} \textbf{99}, 240405 (2007)

\bibitem{Schl01}  W.P. Schleich, Quantum Optics in Phase Space (Wiley, New York, 2001), Chaps. 11 and 16.

\bibitem{Brun92} M. Brune, et al., \emph{Phys. Rev. A} \textbf{45}, 5193 (1992)

\bibitem{Raim01} J.M. Raimond,et al., \emph{Rev. Mod. Phys.} \textbf{73}, 565 (2001)

\bibitem{Ourj06} A. Ourjoumtsev, et al., \emph{Science} \textbf{382}, 83 (2006); A. Ourjoumtsev, et al., \emph{Nature} \textbf{448}, 784 (2007)

\bibitem{DeMa08} F. De Martini, et al., \emph{Phys. Rev. Lett.} \textbf{100}, 253601 (2008).

\bibitem{Scia05} F. Sciarrino, et al., \emph{Phys. Rev. A} \textbf{72}, 062313 (2005).

\bibitem{Scia07} F. Sciarrino, et al., \emph{Phys. Rev. A} \textbf{76}, 012330(2007).

\bibitem{DeMa98} F. De Martini, \emph{Phys. Rev. Lett.} \textbf{81}, 2842
(1998); \emph{Phys. Lett. A} \textbf{250}, 15 (1998).

\bibitem{DeMa05b} F. De Martini et al.,
\emph{Prog. Quant. Elect.} \textbf{29}, 165 (2005).

\bibitem{DeMa05}  F. De Martini, et al.,
\emph{Phys. Rev. Lett.} \textbf{95}, 240401 (2005).

\bibitem{Naga07} E. Nagali, et al.,  \emph{Phys. Rev. A} \textbf{76}, 042126 (2007).

\bibitem{Ricc05} M. Ricci, et al., \emph{Phys. Rev. Lett.} \textbf{95}, 090504 (2005)

\bibitem{DeMa08b} F. De Martini, et al., arXiv:0804.0341.

\bibitem{Zure03} W.H. Zurek, \emph{Phys. Rev. Lett.} \textbf{90}, 120404 (2003).

\bibitem{Jozs94} R. Jozsa, \emph{J. Mod. Opt.} \textbf{41}, 2315 (1994)

\bibitem{Bure69} D. Bures, \emph{Trans. Am. Math. Soc.} \textbf{135}, 199 (1969)

\bibitem{Hubn92} M. Hubner, \emph{Phys. Lett. A} \textbf{163}, 239 (1992); \emph{Phys. Lett. A} \textbf{179}, 226 (1993)

\bibitem{Loud} R. Loudon, The Quantum Theory of Light.

\bibitem{Leon93} U. Leonhardt, \emph{Phys. Rev. A} \textbf{48}, 3265 (1994)

\bibitem{Schl91} W. Schleich, et al., \emph{Phys. Rev. A} \textbf{44}, 2172 (1991)

\bibitem{nota} The closed-form calculations leading to Eq.3 as well as similar expression for $D(x)$ will be reported in a forthcoming paper.

\bibitem{Bjor08} G. Bjork, \emph{private communication}

\bibitem{NOON} J. Jacobson, et al., \emph{Phys. Rev. Lett.} \textbf{74}, 4835 (1995);
K.T. Kapale, et al., \emph{Phys. Rev. Lett.} \textbf{99}, 053602 (2007); F. Sciarrino, et al. \emph{Phys. Rev. A} \textbf{77},012324 (2008).

\bibitem{Slat66} L.J. Slater, \emph{Generalized Hypergeometric Functions}, Cambridge University Press, Cambridge (1966)

\bibitem{Pere95} A. Peres, Quantum Theory: Methods and Concepts (Kluwer Academic Publishers, Dordrecht,1995).

\bibitem{Spag08} N. Spagnolo, et al., \emph{Optics Express} \textbf{16}, 17609 (2008)

\bibitem{Gisi96} B. Huttner, et al., \emph{Phys. Rev. A} \textbf{54}, 3783 (1996)

\end{thebibliography}
\end{document}